\documentstyle{aipproc}

\newcommand{\plb}[2]{{\em Phys. Lett.}          {\bf #1B}, #2 }

\newcommand{\prd}[2]{{\em Phys. Rev.}           {\bf D#1}, #2 }
\newcommand{\prl}[2]{{\em Phys. Rev. Lett.}     {\bf  #1}, #2 }
\newcommand{\zpc}[2]{{\em Z. Phys.}             {\bf C#1}, #2 }
\newcommand{\sci}[2]{{\em Science}              {\bf  #1}, #2 }
\newcommand{\jpb}[2]{{\em J. Phys.}             {\bf B#1}, #2 }

\newcommand{\etal}{{\em et al.}}

\newcommand{\col}{Collaboration}

\newcommand{\be}{\begin{equation}}
\newcommand{\ee}{\end{equation}}
\newcommand{\ba}{\begin{array}}
\newcommand{\ea}{\end{array}}
\newcommand{\ms}{{\overline{\rm MS}}}

\begin{document}
\hspace{320pt}UPR--791--T
\title{Bounds on the Standard Higgs Boson\thanks{To be
published in the proceedings of the Workshop on Physics at the 
First Muon Collider and at the Front End of a Muon Collider, 
Batavia, IL, November 6--9, 1997.}}

\author{Jens Erler and Paul Langacker}

\address{Department of Physics and Astronomy, \\
University of Pennsylvania, \\
Philadelphia, PA 19104-6396, USA}

\maketitle

\begin{abstract}
We review the status of precision electroweak physics with particular
emphasis on the extraction of the Higgs boson mass. Global fit results
depend strongly on the used value for the hadronic contribution to 
$\alpha (M_Z)$. We emphasize, however, that the general tendency
for a light Higgs persists when using any of the recently obtained values
for $\alpha (M_Z)$, and is also less dependent on deviating observables
such as $A_{LR}$ than in the past.
\end{abstract}

Before the discovery of the top quark, precision analyses of the Standard 
Model (SM) were mainly focussed on constraining its mass, $m_t$, while
the Higgs boson mass, $M_H$, was fixed to a set of reference values 
between its direct lower limit and typically 1 TeV. After the top quark was 
discovered~\cite{jerler:Abe95} and its mass found to be in perfect agreement 
with the predictions of precision measurements at LEP and elsewhere, the 
interest shifted towards finding similar constraints for $M_H$. 

With a first precise measurement of the left-right asymmetry, $A_{LR}$, at 
the SLC~\cite{jerler:Abe93} came also for the first time a preference for a 
light Higgs boson from precision tests. Indeed, by changing $M_H$ 
from 1 TeV to 60 GeV the minimum $\chi^2$ decreased by 4.4 units. However, 
this observation depended entirely on the $A_{LR}$ and $R_b$ measurements, 
both of which deviated by more than 2 $\sigma$ from their SM 
predictions. Removing them resulted in a virtually flat $\chi^2 (M_H)$ 
function~\cite{jerler:Erler95}. $R_b$ itself is independent of $M_H$,
but it favors a smaller $m_t$ and through the strong $m_t$--$M_H$
correlation in the $\rho$ parameter, $M_H$ is also driven to smaller values.

Subsequently the $A_{LR}$ and $R_b$ measurements moved closer to the SM,
but with their smaller errors the deviations remained at the 2 $\sigma$
level, and as a result the sensitivity to $M_H$ was enhanced.
The direct top mass determinations by CDF and D\O\ increased the
sensitivity further and the minimum $\chi^2$ value now increased by more 
than 10 when $M_H$ was increased to 1 TeV~\cite{jerler:Langacker96}.
Yet, most of the sensitvity was lost upon removing $A_{LR}$ and $R_b$,
and both, central values and upper limits for $M_H$ depended strongly
on only 2 input quantities, both in conflict with the prediction. 

Extraction of information on $M_H$ is also hampered by the uncertainties in
the hadronic contribution to the vacuum polarization, 
$\Delta \alpha_{\rm had}^{(5)} (M_Z)$. There is a strong (70\%) anticorrelation 
between $\Delta \alpha_{\rm had}^{(5)} (M_Z)$ and $M_H$. 

With the increased precision of the measurements at 
LEP 1~\cite{jerler:Abbaneo97,jerler:Abreu95} and the SLC~\cite{jerler:Abe97}, 
a better agreement of $R_b$ with the SM prediction (1.3 $\sigma$), 
accurate measurements of the $W$ boson mass, $M_W$, at 
LEP 2~\cite{jerler:Abbaneo97} and the 
Tevatron~\cite{jerler:Abe95A,jerler:Abachi96}, and interesting new
developments regarding the determination of 
$\Delta \alpha_{\rm had}^{(5)} (M_Z)$ as we will discuss later, the tendency 
for a light Higgs became stronger. The minimum $\chi^2$ for a 1 TeV Higgs 
boson is now 16.6 larger than at its direct lower limit~\cite{jerler:Erler97}, 
and is less dependent on conflicting observations, although $A_{LR}$ 
continues to play an important role. 

We will now discuss the current status on electroweak precision tests
within the SM. Most of the results presented here are from 
the December 1997 off-year partial update of the Particle Data 
Group (PDG)~\cite{jerler:Erler97} where more details, an extended list 
of references, and constraints on parameters describing physics 
beyond the SM can be found. For implications of electroweak precision 
studies for supersymmetric extensions of the Standard Model see
Ref.~\cite{jerler:Erler97A}. We will conclude with a discussion of the current 
limits on $M_H$, its central fit values for a variety of fits, and the
impact of $\Delta \alpha_{\rm had}^{(5)} (M_Z)$ and some very recent
developments in its determination.

In Table~\ref{T:jerler:1} we give a list of observables used in the fits.
The value of $m_t = 175 \pm 5$~GeV includes results from the dilepton, lepton 
plus jet, and all hadronic channels~\cite{jerler:Abe97A,jerler:Abachi97}. 
$\Gamma_Z$ is the total width 
of the $Z$ boson, $\sigma_{\rm had}$ its hadronic peak cross section, 
and the $R_f$ and $A_{FB}^{(0,f)} = {3\over 4} A_e A_f$ are 
branching ratios (normalized w.r.t.\ the hadronic width)
and forward-backward asymmetries on the $Z$ pole, 
respectively~\cite{jerler:Abbaneo97,jerler:Abreu95}. $A_f$ is a function
of the effective weak mixing angle, $\bar{s}_f^2$, appearing in the $Zff$ 
coupling. The two values of $s^2_W$ from deep-inelastic neutrino 
scattering are from CCFR\cite{jerler:McFarland97} and the global average, 
respectively. Similarly, the $g^{\nu e}_{V,A}$ are from 
CHARM~II\cite{jerler:Vilain94} and from the $\nu e$ scattering world average. 
The second errors in the weak charges, $Q_W$, of atomic parity violation in 
Cs~\cite{jerler:Wood97} and Tl~\cite{jerler:Edwards95} are 
theoretical\cite{jerler:Dzuba97,jerler:Dzuba87}. The value of $\alpha_s$ 
[in brackets] from non-lineshape determinations~\cite{jerler:Hinchliffe97} 
is for comparison only, and is not used as a fit constraint.

\begin{table}
\caption[]{Principal LEP and other recent observables compared with the Standard Model 
predictions for $M_H = M_Z$. The first value for $M_W$ is from $p\bar{p}$ 
colliders~\cite{jerler:Abe95A,jerler:Abachi96}, while the second includes 
the measurements at LEP\cite{jerler:Abbaneo97}. 
The four values of $A_\ell$ are 
(i) from $A_{LR} = A_e$, the left-right asymmetry for hadronic final 
states~\cite{jerler:Abe97}; (ii) the combined value from SLD including 
leptonic asymmetries and assuming univerality;
(iii) $A_\tau$ from the total $\tau$ polarization; and 
(iv) $A_e$ from the angular distribution of the $\tau$ polarization. 
The other $A_f$ are mixed forward-backward left-right asymmetries
from SLD~\cite{jerler:Abbaneo97}. $\bar{s}_\ell^2(A_{FB}^{(0,q)})$ is extracted
from the hadronic charge asymmetry. 
The uncertainties in the SM predictions are from the fit parameters. 
The SM errors in $\Gamma_Z$, $R_\ell$, and $\sigma_{\rm had}$ are completely 
dominated by the uncertainty in $\alpha_s$. In parentheses we show the 
shift in the predictions when $M_H$ is changed to 300~GeV.
Older low-energy results are not listed but are included in the fits.}
\begin{tabular}{ccc}
{\bf Observable}                 & {\bf Value}          & {\bf Standard Model}              \\ \\
\tableline
$m_t$                    [GeV]   &$     175 \pm 5      $&$   173   \pm 4      \;(+5)      $ \\
$M_W$                    [GeV]   &$ 80.405  \pm 0.089  $&$ 80.377  \pm 0.023  \;(-0.036)  $ \\ 
                                 &$ 80.427  \pm 0.075  $&$                                $ \\ 
$M_Z$                    [GeV]   &$ 91.1867 \pm 0.0020 $&$ 91.1867 \pm 0.0020 \;(+0.0001) $ \\ 
$\Gamma_Z$               [GeV]   &$  2.4948 \pm 0.0025 $&$  2.4968 \pm 0.0017 \;(-0.0007) $ \\ 
$\sigma_{\rm had}$       [nb]    &$ 41.486  \pm 0.053  $&$ 41.469  \pm 0.016  \;(-0.005)  $ \\ 
$R_\ell$                         &$ 20.775  \pm 0.027  $&$ 20.754  \pm 0.020  \;(+0.003)  $ \\ 
$R_b$                            &$  0.2170 \pm 0.0009 $&$  0.2158 \pm 0.0001 \;(-0.0002) $ \\ 
$R_c$                            &$  0.1734 \pm 0.0048 $&$  0.1723 \pm 0.0001 \;(+0.0001) $ \\ 
$A_{FB}^{(0,\ell)}$              &$  0.0171 \pm 0.0010 $&$  0.0162 \pm 0.0003 \;(-0.0004) $ \\ 
$A_{FB}^{(0, b  )}$              &$  0.0984 \pm 0.0024 $&$  0.1030 \pm 0.0009 \;(-0.0013) $ \\ 
$A_{FB}^{(0, c  )}$              &$  0.0741 \pm 0.0048 $&$  0.0736 \pm 0.0007 \;(-0.0010) $ \\ 
$A_{FB}^{(0, s  )}$              &$  0.118  \pm 0.018  $&$  0.1031 \pm 0.0009 \;(-0.0013) $ \\ 
$\bar{s}_\ell^2(A_{FB}^{(0,q)})$ &$  0.2322 \pm 0.0010 $&$  0.2315 \pm 0.0002 \;(+0.0002) $ \\ 
$A_\ell$                         &$  0.1550 \pm 0.0034 $&$  0.1469 \pm 0.0013 \;(-0.0018) $ \\ 
                                 &$  0.1547 \pm 0.0032 $&$                                $ \\ 
                                 &$  0.1411 \pm 0.0064 $&$                                $ \\ 
                                 &$  0.1399 \pm 0.0073 $&$                                $ \\ 
$A_b$                            &$  0.900  \pm 0.050  $&$  0.9347 \pm 0.0001 \;(-0.0002) $ \\ 
$A_c$                            &$  0.650  \pm 0.058  $&$  0.6678 \pm 0.0006 \;(-0.0008) $ \\ 
$s^2_W(\nu{\rm N})=1-M_W^2/M_Z^2$&$  0.2236 \pm 0.0041 $&$  0.2230 \pm 0.0004 \;(+0.0007) $ \\ 
                                 &$  0.2260 \pm 0.0039 $&$                                $ \\ 
$g_V^{\nu e}$                    &$ -0.035  \pm 0.017  $&$ -0.0395 \pm 0.0005 \;(+0.0002) $ \\ 
                                 &$ -0.041  \pm 0.015  $&$                                $ \\ 
$g_A^{\nu e}$                    &$ -0.503  \pm 0.017  $&$ -0.5064 \pm 0.0002 \;(+0.0002) $ \\ 
                                 &$ -0.507  \pm 0.014  $&$                                $ \\ 
$Q_W({\rm Cs})$           &$ -72.41 \pm 0.25\pm 0.80   $&$  -73.12 \pm 0.06   \;(+0.01)   $ \\ 
$Q_W({\rm Tl})$           &$ -114.8 \pm 1.2 \pm 3.4    $&$  -116.7 \pm 0.1                $ \\ 
\tableline
$\Delta\alpha_{\rm had}^{(5)}(M_Z)$&$0.02817\pm 0.00062$&$ 0.02802 \pm 0.00049\;(-0.00066)$ \\ 
$\sin^2 \hat\theta_\ms$            &        ---         &$ 0.23124 \pm 0.00017\;(+0.00024)$ \\
$\alpha_s$                         &$[0.1178\pm 0.0023]$&$ 0.1214  \pm 0.0031 \;(+0.0018) $ \\ 
\end{tabular}
\label{T:jerler:1}
\end{table}

If we include $M_H$ as a fit parameter we find
\be
\label{E:jerler:mh}
   M_H = 69^{+85}_{-43} \hbox{ GeV},
\ee
with the central value slightly below the direct lower limit
of 77 GeV (95\% CL)~\cite{jerler:Murray97}. The central value
in Eq.~(\ref{E:jerler:mh}) is 46 GeV smaller
than the best fit value obtained by the LEP Electroweak
Working Group (LEPEWWG)~\cite{jerler:Abbaneo97}. We trace the 
differences to a different treatment of radiative corrections 
and to a slightly different and more recent data set.
Most importantly, inclusion of ${\cal O} (\alpha^2 m_t^2)$
corrections~\cite{jerler:Degrassi96} shift the extracted
$M_H$ by $-17$ GeV. Also, we use the recent update for
$\Delta\alpha_{\rm had}^{(5)}(M_Z)$ from Alemany, Davier, and
H\"ocker~\cite{jerler:Alemany97}, which drives $M_H$ smaller 
by another 10 GeV compared to the use of 
$\Delta\alpha_{\rm had}^{(5)}(M_Z)$ from Eidelman and 
Jegerlehner~\cite{jerler:Eidelman95}. Our result,
$$\alpha_s = 0.1214  \pm 0.0031 \;(+0.0018), $$
is higher than the one in Ref.~\cite{jerler:Abbaneo97}.
This is mainly due to ${\cal O} (\alpha\alpha_s)$ vertex 
corrections~\cite{jerler:Czarnecki96} which increase the
extracted $\alpha_s$ by 0.001. Taking these and other smaller
differences, which are well understood, into account, the agreement 
with the results of the LEPEWWG is excellent. We would 
like to stress that this agreement is quite remarkable
as the electroweak library ZFITTER~\cite{jerler:Bardin92}
is based on the on-shell renormalization scheme, while we use the 
$\ms$ scheme throughout. It also demonstrates
that once the most recent theoretical calculations, in 
particular Refs.~\cite{jerler:Degrassi96,jerler:Czarnecki96}
are taken into account, the theoretical uncertainty becomes 
quite small and is in fact presently negligible compared
to the experimental errors. The relatively large theoretical 
uncertainties obtained in the Electroweak Working Group 
Report~\cite{jerler:Bardin97} were estimated using different
electroweak libraries, which did not include the full range of
higher order contributions available now. 

The agreement between theory and experiment is excellent. Even the largest 
discrepancies in $A_{LR}^0$, $A_{FB}^{(0,b)}$, and $A_{FB}^{(0,\tau)}$, 
deviate by only 2.4~$\sigma$, 1.9~$\sigma$ and 1.7~$\sigma$, respectively.
There is an experimental discrepancy of 1.9 $\sigma$ between $A_\ell$ from
LEP and the SLC,
\be \ba{l}
\label{E:jerler:aell}
  A_\ell ({\rm LEP}) = 0.1461 \pm 0.0033, \\
  A_\ell ({\rm SLD}) = 0.1547 \pm 0.0032,
\ea \ee
where the LEP value is from letponic forward-backward asymmetries and
$\tau$ polarization measurements assuming lepton universality. 
If one considers this discrepancy as a fluctuation, one can use the 
average value from Eqs.~(\ref{E:jerler:aell}) to extract $A_b$ from
$A_{FB}^{(0,b)} = {3\over 4} A_e A_b$ and combine it with $A_b$ from SLD
to obtain $A_b = 0.877 \pm 0.023$,
which is 2.5 $\sigma$ or 6\% below the SM prediction. That means a 
30\% radiative correction to $\hat\kappa_b$ defined through 
$\sin^2 \hat\theta^{\rm eff}_b = \hat\kappa_b \sin^2 \hat\theta_\ms$
would be needed to explain the discrepancy in terms of new physics 
in loops. Only a new type of physics which couples at the tree level
preferentially to the third generation, and which does not contradict $R_b$ 
(including the off-peak $R_b$ measurements by DELPHI\cite{jerler:Abreu96}),
can conceivably account for a low $A_b$\cite{jerler:Erler95A}.

Let us now return to the implication for the Higgs mass. Results
depend strongly on the used input parameter 
$\Delta\alpha_{\rm had}^{(5)}(M_Z)$. There has been a lot of
activity in the recent past on this subject, and initially
not all the obtained results were in agreement with each other. 
This is due to the difficulty of extracting phenomenologically 
the function $R(s)$ describing the cross section for $e^+ e^-$ 
annihilation into hadrons from low and intermediate energy 
collider data. Now, the results obtained from this type of analysis 
are in reasonable agreement. Alternatively, one may try to employ
perturbative QCD (PQCD) down to smaller energies, $\sqrt{s} \sim m_\tau$, 
and compute the continuum contribution to $R(s)$ theoretically. This 
approach was advocated by Martin and Zeppenfeld~\cite{jerler:Zeppenfeld95}, 
and yields both smaller central values and errors for 
$\Delta\alpha_{\rm had}^{(5)}(M_Z)$. The main reason is that some of 
the measured cross sections lie systematically higher than 
the theoretical predictions in a regime where PQCD should be reliable. 
Very recently, Davier and H\"ocker~\cite{jerler:Davier97} improved this 
approach by performing a spectral moment analysis of $R(s)$ and showing 
that the non-perturbative terms are under control 
(and very small). Hence this approach appears to be 
quite reliable. Moreover, a similar technique~\cite{jerler:LeDiberder92} 
applied to $\tau$ decays yields consistent results~\cite{jerler:Hocker97}. 
Therefore, it was concluded in Ref.~\cite{jerler:Davier97} that PQCD
can be applied down to $\sqrt{s} = m_\tau$. 
If we use the resulting $\Delta\alpha_{\rm had}^{(5)}(M_Z) = 0.02784 \pm 0.00026$,
(with the top quark contribution removed) for our fit, we find 
\be
\label{E:jerler:mh1}
  M_H = 93_{-46}^{+76} \hbox{ GeV}.
\ee
Here the central value is above the direct lower limit. 
It should be stressed however, that a precise prediction
for $M_H$ is impossible to obtain due to the large
error, the SLD discrepancy, and the complications from 
$\Delta\alpha_{\rm had}^{(5)}(M_Z)$. On the other hand, 
upper limits and the tendency for a light Higgs are
more robust. The 90 (95)\% upper limits on $M_H$
from the more experimental~\cite{jerler:Alemany97}
and the more theoretical approach~\cite{jerler:Davier97}
are $M_H < 236$ (287) GeV and $M_H < 224$ (266) GeV,
respectively, fortuitously in very good agreement. 
In order to obtain these upper limits, we have taken the
Higgs exclusion curve from LEP~\cite{jerler:Murray97} carefully into
account. Since this curve extends above the quoted lower
limit of 77 GeV, this results in slightly higher (more
conservative) upper limits. 

As a demonstration that the
tendency for a light Higgs is not entirely due to the
high $A_{LR}$ we remove it from the data and the
result~(\ref{E:jerler:mh}) changes to
\be
\label{E:jerler:mh2}
  M_H = 154_{-82}^{+140} \hbox{ GeV}.
\ee
Clearly, the central value and the errors are much larger, but this
result is still compatible with the supersymmetric Higgs mass range
$M_H < 150$ GeV. A less radical way to deal with deviating
data is the use of PDG scale factors. Using them results in an
increase of upper limits by ${\cal O} (100)$ GeV~\cite{jerler:Erler97}.

In conclusion, the SM of electroweak interactions is in excellent
agreement with observations, with only a few deviations
in some asymmetries. There is a much stronger tendency for a
light Higgs boson than in the past, independently of whether
one wishes to rely on PQCD or not. On the other hand, best fit
values for $M_H$ are rather volatile and depend more sensitively
on input parameters and details of the analysis. 

\noindent {\bf Acknowledgement:} We would like to thank D. Zeppenfeld 
for useful discussions and the Aspen Center for Physics for its hospitality.

\end{document}